%% file: main.tex
 \documentclass[journal,twocolumn]{IEEEtran}

\usepackage[utf8]{inputenc}
\usepackage[T1]{fontenc}
\usepackage[
	disable,
backgroundcolor = White,textwidth=\marginparwidth]{todonotes}
\newcommand{\dg}[2][]{\todo[color=red!20,size=\tiny,#1]{DG: #2}} 

\usepackage[markup=underlined]{changes}

\usepackage[utf8]{inputenc} 
\usepackage[T1]{fontenc}    
\usepackage{url}            
\usepackage{booktabs}       
\usepackage{amsfonts}       
\usepackage{nicefrac}       
\usepackage{microtype}      

\usepackage{graphicx}
\usepackage{color}
\usepackage{rotating}
\usepackage{tabularx}
\usepackage{pdflscape}

\usepackage{amsmath,amssymb,amsthm}

\usepackage{blkarray}
\usepackage{algorithm,algorithmic}
\usepackage{bbm,dsfont}
\usepackage{caption,subcaption}
\usepackage{wrapfig}
\usepackage{cancel}
\usepackage{enumerate, cases}
\usepackage{thmtools,thm-restate}
\usepackage{mathtools}

\usepackage[none]{hyphenat}
\usepackage{multirow}
\usepackage{makecell}
\usepackage{setspace}
\allowdisplaybreaks

\usepackage{dblfloatfix}
\usepackage{array}

\usepackage{enumitem}

\allowdisplaybreaks

\DeclareMathOperator*{\argmax}{arg\,max}

\newcolumntype{C}[1]{>{\centering}m{#1}}
\newcommand{\EE}[1]{\mathbb{E}\left[#1\right]}

\newcommand{\Prob}[1]{\mathbb{P}\left\{#1\right\}}

\newcommand{\Regret}{\mathcal{R}}

\newcommand{\R}{\mathbb{R}}

\usepackage{graphicx}
\usepackage{mathptmx}
\usepackage{amsmath}
\usepackage{amssymb}
\usepackage{amsthm}

\newtheorem{thm}{Theorem}
\newtheorem{lem}{Lemma}

\theoremstyle{definition}

\usepackage{verbatim}

\begin{document}
\title{\huge \bf
Learning to Optimize Energy Efficiency in Energy Harvesting Wireless Sensor Networks 
}


\author{Debamita Ghosh, Manjesh K. Hanawal and Nikola Zlatanov
\thanks{Debamita Ghosh is Ph.D. student in the Dept. of IEOR, IITB-Monash Research Academy, IIT Bombay, India.}
\thanks{Manjesh K. Hanawal is with the Dept. of IEOR, IIT Bombay, India.}  
\thanks{Nikola Zlatanov is with the Dept. of ECSE, Monash University, Australia.}
}
\maketitle

\begin{abstract}

We study wireless power transmission by an energy source to multiple energy harvesting nodes with the aim to maximize the energy efficiency. The source transmits energy to the nodes using one of the available power levels in each time slot and the nodes transmit information back to the energy source using the harvested energy. The source does not have any channel state information and it only knows whether a received codeword from a given node was successfully decoded or not. With this limited information, the source has to learn the optimal power level that maximizes the energy efficiency of the network. We model the problem as a stochastic Multi-Armed Bandits problem and develop an Upper Confidence Bound based algorithm, which learns the optimal transmit power of the energy source that maximizes the energy efficiency. Numerical results validate the performance guarantees of the proposed algorithm and show significant gains compared to the benchmark schemes. 

\end{abstract}

\begin{IEEEkeywords}
		Multi-Armed Bandits, Energy Harvesting, Wirelessly Powered Communication Network, Energy Efficiency
		 \vspace{-2mm}
\end{IEEEkeywords}

	
\section{Introduction}
\label{sec:introduction}
\input{introduction}

\section{System Model}
\label{sec:system_model}
\input{system_model}

\section{Problem Formulation}
\label{sec:problem_formulation}
\input{problem_formulation}

\section{Numerical Results}
\label{sec:experiment}
\input{experiment}
\section{Conclusion}
\label{sec:conclusion}
\input{conclusion}

\section*{Appendix A}
\label{sec:appendix A}
\input{Appendix}





\bibliographystyle{IEEEtran}
\bibliography{ref} 

\end{document}

%% file: introduction.tex
Wirelessly powered communication networks (WPCNs), where nodes can perform Energy Harvesting (EH), have recently drawn an upsurge in interest as a possible technology for next generation wireless networks. EH helps to enhance the life of wireless nodes by continuously charging their batteries either by scavenging energy from the ambient atmosphere or from a dedicated energy source. When a dedicated energy source is used for energy transmission, the amount of energy it emits has to be appropriately calibrated. In the one hand, high-power emissions may result in energy wastage due to the fundamental limitations of the amount of energy EH devices can harvest. On the other hand, low-power emissions may result in EH devices not being able to harvest enough energy. 
It is thus important that the energy source transmits with the right amount of power. As a result, the metric Energy Efficiency (EE), measured in bits-per-joule, has drawn considerable attention as an essential performance metric for optimizing the transmit power in WPCNs. Motivated by this, our goal in this paper is to develop an algorithm that learns the optimal power level of an energy source that maximizes the EE in WPCNs. 

Several authors have investigated EE of various WPCNs. For example, the authors in \cite{shi2015energy} study the maximization of EE in multi-users multiple-input-single-output systems with Simultaneous Wireless Information and Power Transfer (SWIPT). The authors in \cite{ng2013wireless} study EE of Orthogonal Frequency Division Multiple Acesss (OFDMA) systems employing SWIPT. The authors in \cite{wu2016user}, maximize the weighted sum of the user's EEs by jointly optimizing the used power for downlink wireless energy transmission (WET) and for uplink wireless information transmission (WIT). Similar joint transmitter and receiver optimization of EE of OFDMA systems taking into account the effects of active subcarriers and users is studied in \cite{wu2014resource}. The authors in \cite{guo2017energy} study the EE of SWIPT in mobile wireless sensor networks considering the minimum individual and system data rate requirements of the receivers, the minimum required power transfer, and the maximum system power consumption. 

The works \cite{shi2015energy}-\cite{guo2017energy}, as well as others in the literature, study EE in WPCNs assuming that the channel state information (CSI) is known at the receivers and/or the transmitters. However, acquiring CSI consumes energy, which diminishes the EE of the system. As a result, we propose a learning algorithm that maximizes the EE of the system without any CSI knowledge. Specifically, we consider a WPCN where the energy source transmits energy to EH nodes and the nodes use the harvested energy to transmit information back to the energy source at a fixed rate. The energy source then tries to decode the received codewords and observes which nodes were successful in their transmission. Using this limited information regarding the successful decoding of received codewords, the source learns which power level to use in order to maximise the EE of the system. 
To this end, we model the problem as a stochastic multi-armed bandits problem and 
develop a learning algorithm to optimize EE. 

Learning algorithms in WPCN were investigated recently in \cite{luo2019learning} and \cite{min2019learning}. The authors in \cite{luo2019learning} propose reinforcement learning (RL) to investigate the tradeoff between EE and quality of service parameters, such as transmission delay, in device-to-device communications underlaying cellular networks. The authors in \cite{min2019learning} propose a RL based offloading policy for EH IoT devices to select the edge devices for computationally intensive applications. These learning algorithms require that the devices either have CSI or know additional information such as the available battery power, tasks which are hard to realize in practice and also that diminish the EE. n this paper, we adopt the multi-armed bandit framework for solving the EE problem since it enables us to establish theoretical guarantees on the performance of our algorithm. Thereby, we propose an Upper Confidence Bound (UCB) based learning algorithm using which the energy source maximizes the EE without any CSI and only based on whether a received codeword was successfully decoded or not. To the best of our knowledge, maximizing the EE of a WPCN by optimizing the power selection strategy without CSI knowledge has not been studied previously. 


%% file: system_model.tex
\begin{figure}[!h]
	\centering
	\includegraphics[width=9cm, height = 4cm]{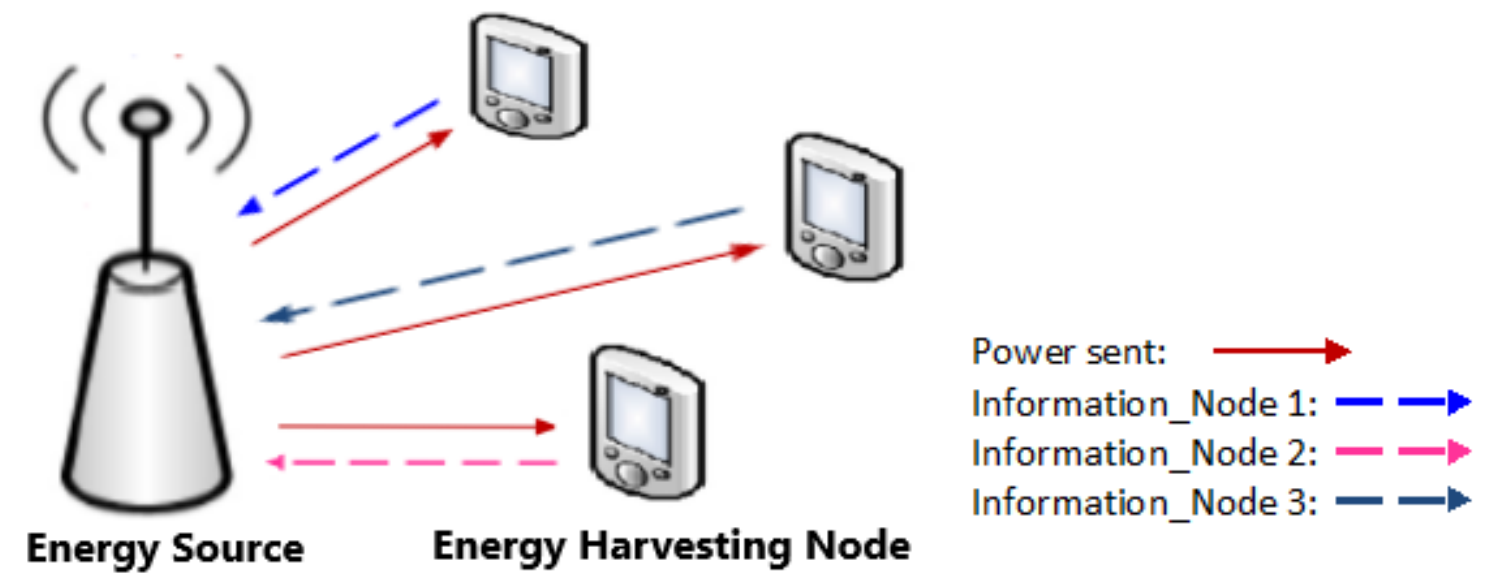}
	\caption{\small{Energy Harvesting setup. Refer \cite{CST2014_wirelessnetwork_LuXiaoPingDusit} for a detailed architecture.}}
	\label{fig:Model}
	\vspace{-4mm}
\end{figure}

We consider a system model comprised of an energy source, referred to as the source, and $k$  EH wireless nodes, referred to as the nodes. The source and the nodes employ frequency-division duplex (FDD)\footnote{Alternatively, we can also assume that the source and the nodes employ time-division duplex (TDD) on $k+1$ time slots in order to avoid interference.} on $k+1$ dedicated channels where each channel occupies a different frequency band. Specifically, the source radiates energy on its dedicated channel, which is then harvested by the nodes and used to transmit information back to the source on their respective dedicated channels, cf. Fig. \ref{fig:Model}. Note that since FDD is used, interference does not occur. The source also serves as a receiver of information. We assume that the dedicated channels are flat-fading channels with channel gains that are constant in each time slot and change independently from one time slot to the next. Let $G_{j}(t)$ denote the complex-valued channel gain between the source and node $j$ in time slot $t$ on the channel dedicated for energy transmission, where $G_{j}(t)\sim\mathbb{CN}\{0,\sigma_{G_j}^2\}$, i.e., complex Gaussian distributed with mean zero and variance $\sigma^2_{G_j}$. Let $H_j(t)$ denote the complex-valued channel gain between node $j$ and the source in time slot $t$ on the channel dedicated for information transmission by node $j$, where $H_j(t)\sim\mathbb{CN}\{0,\sigma_{H_j}^2\}$.  
We also assume that the received signal at the source is corrupted by complex zero-mean additive white Gaussian noise (AWGN) with variance $\sigma^2$.

The source has a set of $m$ powers\footnote{The proposed algorithm works also with a continuum of power,  provided we quantize the continuous power into discrete power. Then, the proposed algorithm will learn the optimal discrete power that is close to the optimal continuous power.} denoted by  $\mathcal{P}=\{p_1, p_2,\dots,p_m\}$. In each time slot $t$, the source can select any power in $\mathcal{P}$ to transmit energy. Let $P(t) \in \mathcal{P}$ denote the power selected in time slot $t$. The amount of energy harvested by node $j$ in time slot $t$ from power $P(t)$, denoted by $E_{j}(t)$, is given by
\begin{align}
\label{eqn:energy harvested}
    E_{j}(t)=\min\{b_{\max},\max\left(0,\lambda P(t) |G_{j}(t)|^2 - p_{\min}\right)\},
\end{align}
where $p_{\min}$ is the minimum required power for each node to operate, $0 \leq \lambda <1$ is the EH inefficiency coefficient, and $b_{\max}$ is the maximum capacity of the battery of the nodes. We assume that the nodes use the entire energy harvested in time slot $t-1$ for information transmission in the next time slot $t$. Each node transmits information using a capacity-achieving codeword with rate $r_0$. 
The received rate at the source from node $j$ in time slot $t$ is denoted by $R_j(t)$ and is given by
\begin{align}
\label{eqn: received rate}
    R_{j}(t) & =O_{j}(t) r_0,
\end{align}
where $O_j(t)$ is a binary variable that models success or failure of the decoding of the codeword of node $j$ at the source, given as
\begin{align}\label{eq_O}
    O_{j}(t) &=\left\{\begin{array}{ll}
        1      & \textrm{if } \log_2\left(1+\frac{E_{j}(t-1) |H_j(t)|^2}{\sigma^2}\right)> r_0  \\
        0      & \textrm{otherwise}.  
 \end{array}
 \right.
 \end{align}
Using $R_j(t),\forall j$, we can define the weighted sum rate received at the source in time slot $t$ as
\begin{align}
R_{sum}(t) =\sum\limits_{j=1}^{k}\omega_jR_j(t),\label{eqn:sum_data_rate}
\end{align}
where $0 \leq \omega_j \leq 1, \forall j$, and $\sum_{j=1}^k\omega_j = 1$. The weights $\omega_j, \forall j$, are used to give different priorities to the users. 


%% file: problem_formulation.tex
In the following, we formulate the problem and develop a learning algorithm to maximize the EE in the WPCN.
\vspace{-2mm}
\subsection{Problem Setting}
\label{sec:problem_setting}
\input{problem_setting}
	
\subsection{Notation}
\label{sec:solution}
\input{solution}

\subsection{The Proposed Algorithm}
\label{sec:algorithm}
\input{algorithm}

%% file: problem_setting.tex
As we increase the transmit power of the source, the amount of energy harvested by the nodes increases only to a certain point and then saturates due to  \eqref{eqn:energy harvested}. As a result, further increase in the transmit power of the source will not increase the weighted sum rate received at the source given by \eqref{eqn:sum_data_rate}, thereby leading to a wastage of energy. Hence, there must be some optimal power level which maximizes the weighted sum rate per energy spent. 
To investigate this point, we define the EE as a ratio between the weighted sum rate received at the source and source's transmit power over $n$ time slots given as
\begin{align}
 EE(n) = \frac{1}{n}\sum_{t=1}^n\sum_{j=1}^k \frac{\omega_jR_j(t+1)}{P(t)}.
\end{align} 
Our goal is to identify a fixed power level that maximizes the expected EE, i.e., we solve the following optimization problem
\begin{align}
\arg\max\limits_{p \in \mathcal{P}}\EE{EE(n) \mid P(t) = p, \; \forall \; t}\label{eqn:NTP}.
\end{align}
 The rationale for seeking a fixed power level in all time slots is because the source does not have CSI and hence adapting the power levels in each time slot is not useful. The only information available to the source is whether a received codeword from a given node was successfully decoded or not. 
Based on this limited information, we derive a strategy for the source to learn the optimal power that maximizes the expected value of the EE defined by \eqref{eqn:NTP}. 
\vspace{-3mm}

%% file: solution.tex
We refer to each power in the set $\mathcal{P}$ simply by its index.  Let $\boldsymbol{R}(t)=\{R_1(t), R_2(t),...,R_{k}(t)\}$ denote the vector of received rates at the source in time slot $t$, where $R_j(t)$ is the received rate from node $j$, given by \eqref{eqn: received rate}. Note that the vector $\boldsymbol{R}(t)$ depends on the energy harvested by the nodes in the previous time slot $t-1$. At $t=1$, $R_j(1)=0, \forall j$, as the nodes do not have any energy harvested from $t=0$. 
Let the history at the source in time slot $t$ be defined as $\mathcal{H}(t):= \left\{(P(1),\boldsymbol{R}(1)),(P(2),\boldsymbol{R}(2)),..,(P(t),\boldsymbol{R}(t))\right\}$ with $\mathcal{H}(0) = \phi$. The power selection policy of the source denoted by $\Pi = \left\{\pi(t)\right\}_{t=1}^{\infty}$ is a sequence of maps $\pi(t):\mathcal{H}(t-1) \rightarrow \mathcal{P}$. 
Our goal is to find a policy that maximizes the expected EE of the system given by \eqref{eqn:NTP}. To this end, we set up the problem as a Multi-Armed Bandits (MAB) problem, where the source is the learner and the set of transmit powers at the source constitute the arms.

Let $R_{j}^i(t)$ denote the rate received at the source from node $j$ in time slot $t$, when the transmit power of the source in time slot $t-1$ was $P(t-1)=i$. For each node $j$ and power $i$,  the random process $\left\{R_j^i(t)\right\}_{t\geq 2}$ is an independent and identically distributed (i.i.d.) process with a mean denoted by $\mu_{ij}$ and given by $\EE{R_j(t+1)|P(t)=i} = \mu_{ij}$. We denote the power that maximizes the expected EE by $i^{*}$ and denote the corresponding optimal expected EE by $\mu_{i^{*}}$, where 
\begin{align}
    i^{*} = \argmax_{i\in m}\sum_{j=1}^{k}\frac{\omega_j\mu_{ij}}{p_i} \quad \text{and}\quad \mu_{i^{*}} = \sum_{j=1}^{k} \frac{\omega_j\mu_{i^*j}}{p_{i^*}}.
\end{align}
We define the expected regret that evaluates the performance gap between the proposed strategy and the optimal power selection strategy on average over $n$ time slots as $\Regret_{n} = n\mu_{i^{*}} - \sum_{t=1}^{n} \sum_{j=1}^{k}\omega_j\EE{\frac{\mu_{I_{t}j}}{p_{I_{t}}}}$,
where $I_t$ is the power index selected in time slot $t$ and $p_{I_t}=P(t)$ is the associated transmit power. 
Hence, maximizing the average EE is equivalent to designing a learning algorithm that minimises the expected regret by learning the optimal power level at the source.

\vspace{-2mm}

%% file: algorithm.tex
We develop an algorithm for optimal power selection using the well known UCB approach \cite{lattimore2020bandit}. We begin by giving a concentration bound on the sum of weighted random variables that will be used in the algorithm.  Let $\hat\mu_{ij}^{(s)}$ denote the empirical mean obtained from $s$ i.i.d. samples of the received rate at the source from node $j$ using the energy harvested from power $i$, given by $\hat\mu_{ij}^{(s)}=\frac{1}{s}\sum_{t=1}^sR_j^i(t)$.
Since $R_j^i(t)\in \{0,r_0\}$, by Hoeffding's Lemma \cite[eq. (2.2)]{book2012_bubeck_regret} there exits a convex function $\psi:\mathbb{R}\rightarrow \mathbb{R}_+$ such that for all $a > 0$,
 \begin{align}
  \EE{\exp(a(\mu_{ij}-R_j^i))} \leq \exp(\psi(a)),\label{eqn:Hoeffding lemma}
 \end{align}
where the convex function $\psi(a)=a^2r_0^2/8$. 

Using \eqref{eqn:Hoeffding lemma} we get the following concentration bound on the weighted sum rate received at the source.
\begin{lem}
 \label{lem:ConcentrationBound1}
 For any $s$, we have 
 \begin{align}
\Prob{\sum\limits_{j=1}^{k}\omega_j\mu_{ij} - \sum\limits_{j=1}^{k}\omega_j\hat{\mu}_{ij}^{(s)}> \epsilon} \leq \exp(-s\psi^*(\epsilon)),\nonumber
\end{align}
 where $\psi^*(\epsilon)=\sup_{a \in \mathbb{R}} (\alpha\epsilon - \sum_{j=1}^{k}\omega_j^2\psi(a))$ is the (weighted) Legendre-Fenchel transform of $\psi(\alpha)$.
\end{lem}
\textit{Proof:} Please refer to Appendix A.

Lemma \ref{lem:ConcentrationBound1} states that with probability $1-\delta$, $\sum_{j=1}^k\omega_j\hat{\mu}^{(s)}_{ij} + (\psi^*)^{-1}\left(\frac{1}{s}\log \frac{1}{\delta}\right) > \sum_{j=1}^k\omega_j\mu_{ij}$ holds $\forall i\in m.$ 
Using this lemma we define UCB for each power $i$ in time slot $t > m$ as
\begin{align}
\label{eqn:UCB2}
\mbox{UCB}_{i}(t)=\sum\limits_{j=1}^k\omega_j\hat\mu_{ij}(t-1) + (\psi^*)^{-1}\left (\frac{\alpha \ln t}{N_{i}(t-1)}\right),
\end{align}
where $\alpha$ is an input parameter to the algorithm, $N_i(t) = \sum_{s=1}^{t}\mathbb{I}\{I_s = i\}$ denotes the number of times the power $i$ is selected till time slot $t$ and $(\psi^{*})^{-1}\left(\frac{\alpha \ln t}{N_{i}(t-1)}\right)= r_{0}\sqrt{\frac{\alpha \ln t\sum_{j=1}^{k}\omega_j^2}{2N_{i}(t-1)}}$. The larger the value of $\alpha$, more is the exploration. We propose a UCB based algorithm whose pseudo-code is given in Algorithm: \ref{alg:UCB}, which 
works as follows. The \ref{alg:UCB} algorithm takes $m$ powers, $k$ nodes, and $\alpha$ as inputs, where $\alpha$ is a constant to be specified later. In the first $m$ time slots, each power is selected in a round-robin fashion. In each of the subsequent time slot $t$, the UCB value of power $i$ is calculated as per (\ref{eqn:UCB2}). The source then selects the power index $i$ which maximizes the ratio $\text{UCB}_i(t)/p_i$. The maximizing power index is denoted by $I_t$ and the transmit power by $p_{I_t}$. For the selected $I_t$, the source then observes the received rates $R_j^{I_t}(t)$ and updates the empirical means $\hat{\mu}_{I_tj}(t)$ for all of the nodes.
\vspace{-2mm}
\begin{algorithm}[H] 
	\renewcommand{\thealgorithm}{UCB-EH}
	\floatname{algorithm}{}
	\caption{\bf }
	\label{alg:UCB}
	\begin{algorithmic}[1]
		\STATE \textbf{Input:} $m, k, \alpha \text{ and } \omega_j, \forall j$
		\STATE Select each transmit power once in first $m$ time slots
		\STATE Update $\hat{\mu}_{ij}$ for all $i\in m, j\in k$
		\FOR{$t= m+1, m+2,\ldots, n$}
			\STATE For each power $i\in m$ calculate 
				\begin{align*}
			\text{UCB}_{i}(t)  \leftarrow \sum\limits_{j=1}^{k}\omega_j\hat\mu_{ij}(t-1) + r_{0}\sqrt{\frac{\alpha \ln t\sum_{j=1}^{k}\omega_j^2}{2N_{i}(t-1)}}	
			\end{align*}
			\STATE Set $I_{t} \leftarrow \argmax\limits_{i}\frac{\text{UCB}_{i}(t)}{p_i}$ and transmit power $p_{I_t}$
            \STATE Observe  received rate $R_{j}^{I_t}(t)$ for all $j\in k$ 
            \STATE  Update the estimates of $\hat{\mu}_{I_tj}(t)$ for all $j \in k$
		\ENDFOR
	\end{algorithmic}
\end{algorithm}
\vspace{-6mm}
\subsection{Regret Analysis}
Now we give a theoretical guarantee on the performance of the proposed \ref{alg:UCB} algorithm. 
\begin{thm}
    \label{thm: Theorem 1}
    Let $\Delta_{i} = \mu_{i^{*}}- \sum_{j=1}^{k}(\omega_j\mu_{ij}/p_{i})$ be the sub-optimality gap of power $i$ and $\Delta = \min\limits_{i \neq i^{*}}\Delta_{i}$. Then, the regret of \ref{alg:UCB} after $n$ rounds is upper bounded as
	\begin{align}
	\label{regret:Thm1}
	    \hspace{-3mm}\Regret_{n} &\leq 6r_{o}^2\sum\limits_{i: \Delta_{i}>0}\frac{\ln n\sum_{j=1}^{k}\omega_j^2}{p_i^2\Delta_{i}} + \sum\limits_{i: \Delta_{i}>0}\left(\frac{\pi^{2}}{3} + 1\right)\Delta_{i},
	 \end{align}
i.e., the regret is of the order $O\left(\frac{m \ln n\sum_{j=1}^{k}\omega_j^2}{\Delta}\right)$.
Further, the problem independent bound is given as $\Regret_{n} \leq O\left(\sqrt{nm\ln  n\sum_{j=1}^{k}\omega_j^2}\right).$
\end{thm}
\textit{Proof:} Please refer to Appendix A.

As seen from Theorem \ref{thm: Theorem 1}, the proposed UCB-EH algorithm is asymptotically optimal and thereby maximizes the expected EE. Moreover, the complexity of the proposed algorithm grows linearly with the number of power levels $m$.

%% file: experiment.tex
Assuming Rayleigh fading, 
$|G_j(t-1)|^2$ and $|H_j(t)|^2$  in $E_j(t-1)$ in \eqref{eqn:energy harvested} are exponentially distributed RVs with means $2 \sigma_{G_j}^2$ and $2 \sigma_{H_j}^2$, respectively. The variances $\sigma_{G_j}^2$ and $\sigma_{H_j}^2$ are obtained using the free-space path loss model as $\sigma_{x_j}^2=\frac{1}{2} 
\left(\frac{c}{4\pi f_{x_j}}\right)^2 d^{-\gamma}_j, x \in\{G,H\}$,
where  $c$ denotes the speed of light, $f_{x_j}$ for $x \in\{G,H\}$ is the carrier frequency of the signal to/from the node $j$, $d_j$ is the length of the link between the source and node $j$, and $\gamma$ is the path loss exponent. Moreover, the noise power $\sigma^2$ in \eqref{eq_O} is given by $\sigma^2=W \sigma_0^2$, where $W$ is the bandwidth and $\sigma_0^2$ is the noise power per Hz.   
For our numerical simulations, we assume a bandwidth $W=100$ kHz and noise power per Hz  $\sigma^2_0=-170$ dBm/Hz, that makes the total noise power $\sigma^2=-120$
dBm. We assume carrier frequencies as $f_{G_j}=2.4$ GHz and $f_{H_j}=2.4+(10^{-3}\times j)$ GHz, the length $d_j=10+(3\times j)$ m, $\omega_j = 1/k$ for $j=1,2,..k$. Furthermore, we consider the path loss exponent $\gamma$ = 2.5, 
the EH inefficiency coefficient $\lambda=0.5$, the minimum power $p_{\min}=-60$
dBm, the maximum battery capacity $b_{\max} \geq -40$ dBm,  and $\alpha=3$. 
For our EH setup, we consider the set of transmit powers as $\mathcal{P} = \{0,1,2,\dots,30\}$ in dBm. 

We empirically evaluate the performance of our proposed algorithm \ref{alg:UCB} in terms of average EE and compare it with 
a benchmark scheme without CSI. In the benchmark scheme without CSI the source always transmits with the maximum power to charge the EH nodes, i.e., it transmits with power $30$ dBm. We define the oracle scheme as the scheme based on the optimal policy which maximizes the energy efficiency of the system. The aim is our algorithm to learn the oracle scheme with time. 
We measure the performance of our algorithm for $T = 10^4$ time slots and the results are averaged over $1000$ repetitions. 

\begin{figure}[H]
\centering
	\includegraphics[width=9.5cm, height = 4.3cm]{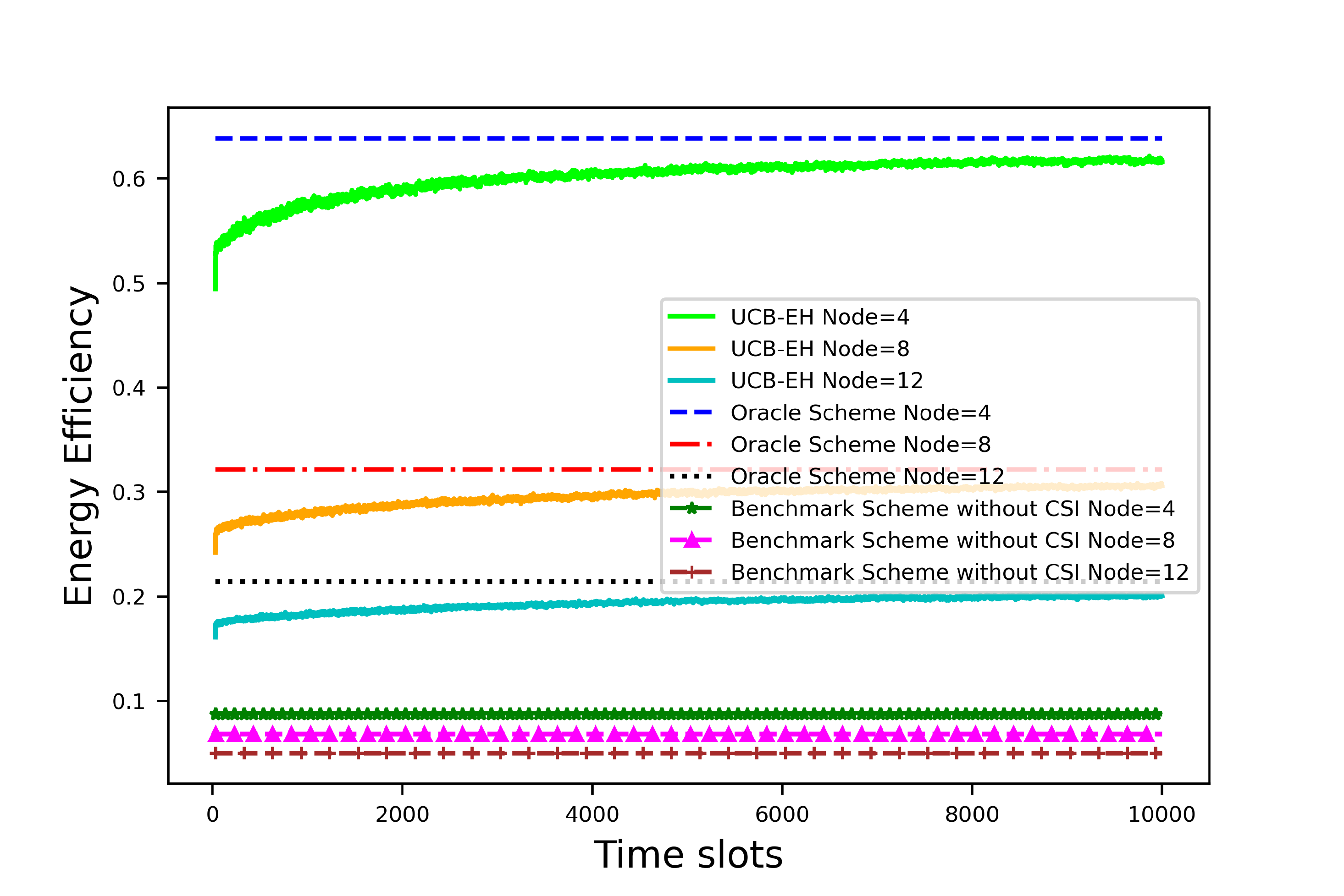}
	\caption{Energy Efficiency v/s Time Slots.}
	\label{fig:ee_vs_nodes}
\vspace{-2mm}
\end{figure}

\begin{figure}[H]
\centering
		\includegraphics[width=9.5cm, height = 3.8cm]{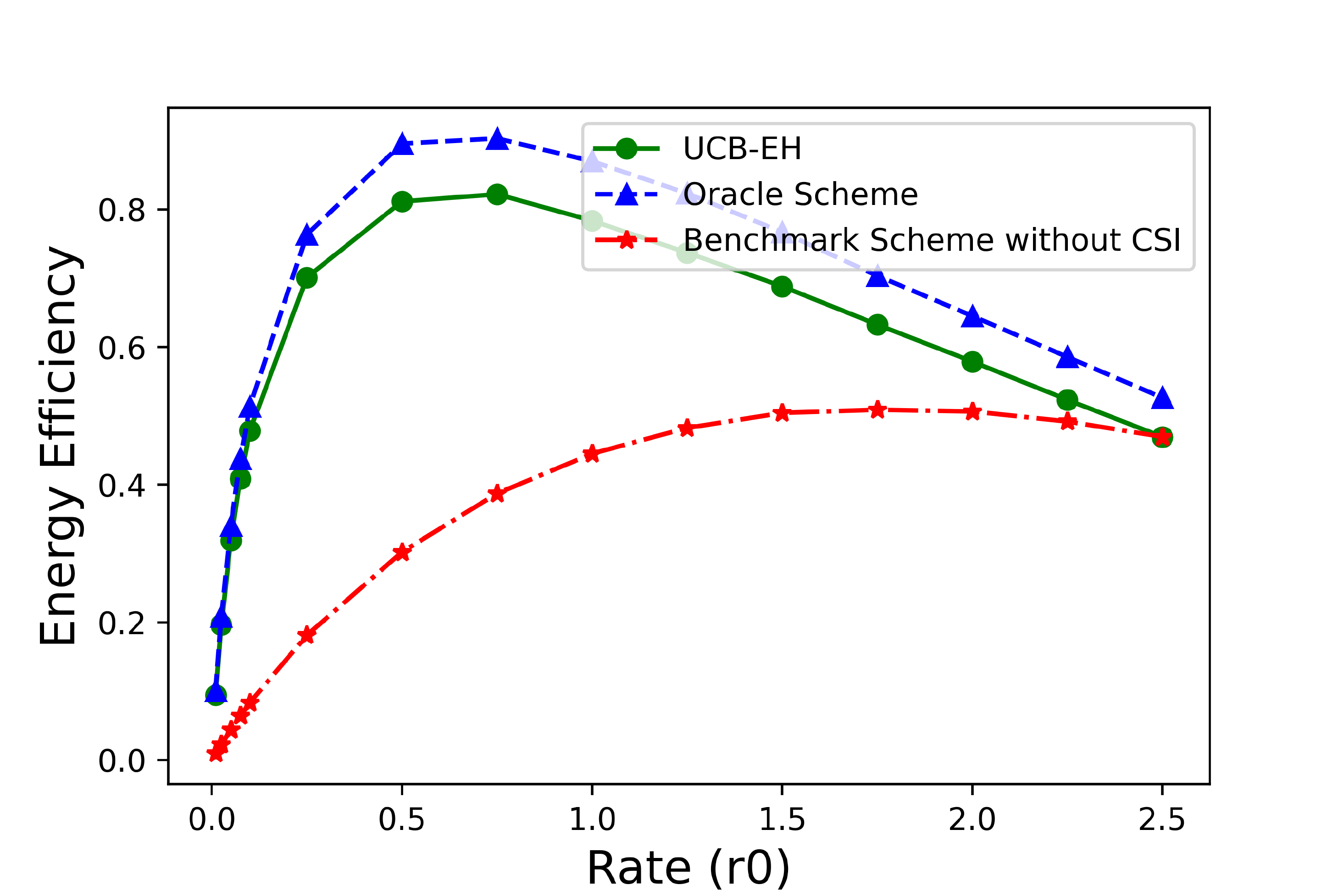}
    	\caption{Energy Efficiency v/s Rate $r_0$ (in bpcu).}
      	\label{fig:ee_r0}
\vspace{-2mm}
\end{figure}

\begin{figure}[H]
	\centering
	\includegraphics[width=9.5cm, height = 3.6cm]{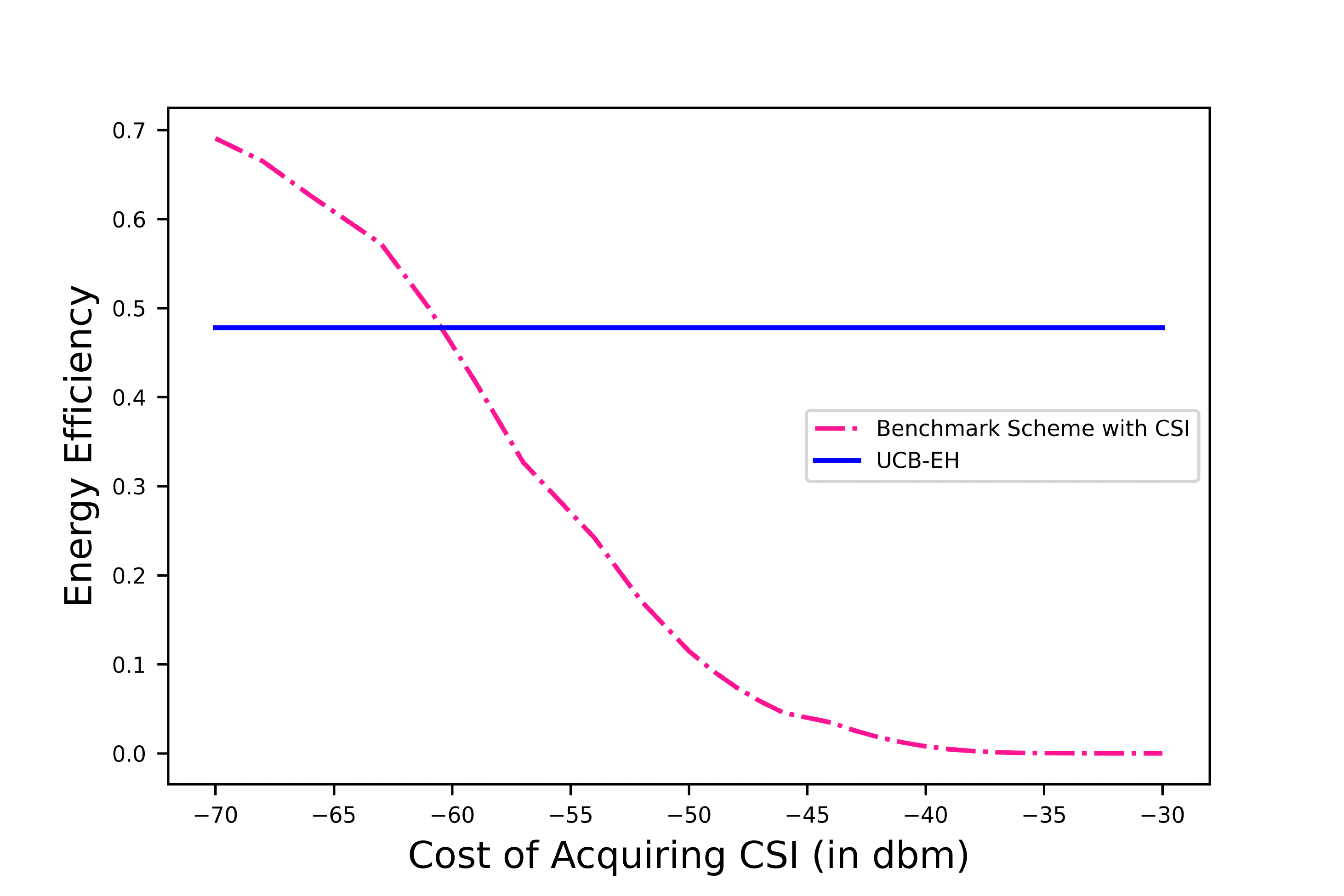}
	\caption{\small{Energy Efficiency with and without CSI.}}
	\label{fig:ee_oracle}
\end{figure}

In Fig. \ref{fig:ee_vs_nodes}, we show the EE achieved with the proposed algorithm, the oracle scheme, and the benchmark scheme without CSI for different numbers of nodes $k =\{4,8,12\}$ and $r_0 = 0.1$ bpcu as a function of the number of time slots $n$. We find that the EE decreases as the number of nodes increases for a fixed number of transmit powers, as expected. As Fig. \ref{fig:ee_vs_nodes} shows, our algorithm achieves an EE that converges toward the EE achieved with the oracle. 
Moreover, our algorithm significantly outperforms the benchmark scheme without CSI for any $k$ and for any time slot.
\vspace{-4mm}
In Fig. \ref{fig:ee_r0}, we show the EE achieved with the proposed algorithm, the oracle scheme, and the benchmark scheme without CSI as a function of the rate $r_0$ when we have $k=5$ nodes. Initially, as we increase the value of $r_0$, more number of nodes can successfully transmit a codeword with rate $r_0$ to the source for a fixed transmit  power level at the source. As a result, the expected EE of the system increases and reaches its maximum point. But as $r_0$ increases further, more number of nodes fail to successfully transmit a codeword with rate $r_0$ to the source which will decrease the EE of the system. As Fig. \ref{fig:ee_r0} shows, our algorithm achieves an EE which is $52\%$ higher than the benchmark scheme without CSI and just $9\%$ lower than the oracle scheme. Moreover, to get the system's highest EE, our algorithm and the oracle scheme need a rate of $0.75$ bpcu, while the benchmark scheme without CSI requires $2$ bpcu to get the system's highest EE.
\vspace{-4mm}
In Fig. \ref{fig:ee_oracle}, we compare the EE of our scheme   with a benchmark scheme that has CSI knowledge of all channels for $k=8$. For a fair comparison, we have included in the  benchmark scheme  the cost of acquiring   CSI.  As can be seen from Fig. \ref{fig:ee_oracle}, the benchmark scheme with CSI  outperforms our algorithm only if the CSI cost is very small, i.e., less than -60 dBm in this example, which is not possible in practice since CSI estimation consumes a fair bit of energy.
\vspace{-4mm}
\vspace{-2mm}

\vspace{-3mm}

%% file: conclusion.tex
We investigated the problem of optimal power selection at a dedicated source that wirelessly charges EH nodes in a 
WPCN, for improving the EE of the system without any CSI knowledge. To address this problem, we proposed an online learning approach based on the theory of multi-armed bandit that selects the optimal transmit power which maximizes the EE of the system. 
The simulation results shows that the proposed algorithm learns the optimal power and it significantly outperforms the benchmark schemes. 

%% file: Appendix.tex
\begin{proof}[\bf Proof of Lemma \ref{lem:ConcentrationBound1}]
Let $\hat\mu_{ij}^{(s)}$ denote the empirical mean based on $s$ i.i.d. samples of node $j$ using the energy harvested from power $i$. Using Markov’s inequality for some $a>0$, we obtain 
\begin{align}
    &\Prob{\sum\limits_{j=1}^{k}\omega_j\mu_{ij} - \sum\limits_{j=1}^{k}\omega_j\hat{\mu}_{ij}^{(s)}> \epsilon} \nonumber\\
     &=e^{-(as\epsilon)}\prod_{j=1}^{k} \prod _{t=1}^s \EE{\exp\left\{a\omega_j(\mu_{i,j}-R^i_j(t))\right\}} \label{eqn:f1_lemma}\\
    &\leq e^{-(a s\epsilon)}\prod_{j=1}^{k} \prod _{t=1}^s \exp\{\psi(a \omega_j)\} \leq \exp\{-s\psi^*(\epsilon)\}.\label{eqn:f2_lemma}
\end{align}
As $\R_j^i(t)$ are i.i.d. across time slots and nodes, we obtain \eqref{eqn:f1_lemma}. We insert \eqref{eqn:Hoeffding lemma} and $\psi(a\omega_j)= \omega_j^2\psi(a)$ in \eqref{eqn:f1_lemma} to get \eqref{eqn:f2_lemma}. \qedhere
\end{proof}
\begin{proof}[\bf Proof of Theorem \ref{thm: Theorem 1}]
 Without loss of generality, let  $i^{*}=1$ be the optimal index of the power. Hence, the optimal power is $p_{1}$. Recall that by convention $\Delta_{i}=\sum_{j=1}^{k}\omega_j\left(\frac{\mu_{1j}}{p_{1}} - \frac{\mu_{ij}}{p_{i}}\right)$. If $I_t = i$ is a suboptimal power selected in time slot $t$, then at least one of the following equations must be true
\begin{align}
    &\sum\limits_{j=1}^{k}\omega_j\hat\mu_{1j}(t-1) + r_{0}\sqrt{\frac{\alpha \ln t\sum_{j=1}^{k}\omega_j^2}{2N_{i}(t-1)}} \leq \sum\limits_{j=1}^{k}\omega_j \mu_{1j}, \label{eqn:UnderEstimate} \\
    &\sum\limits_{j=1}^{k}\omega_j\hat\mu_{ij}(t-1) - r_{0}\sqrt{\frac{\alpha \ln t\sum_{j=1}^{k}\omega_j^2}{2N_{i}(t-1)}} \geq \sum\limits_{j=1}^{k}\omega_j \mu_{ij}, \label{eqn:OverEstimate}\\
    & \qquad \qquad \qquad \Delta_{i} < \frac{2r_{0}}{p_i}\sqrt{\frac{\alpha \ln n\sum_{j=1}^{k}\omega_j^2}{2N_{i}(t-1)}}. \label{eqn:GapBound}
\end{align}
Let us assume that all three conditions are false. Then, we have
\begin{subequations}
\begin{align}
     &\sum\limits_{j=1}^{k}\frac{\omega_j\hat\mu_{1j}(t-1)}{p_1} + \frac{r_0}{p_1}\sqrt{\frac{\alpha \ln t\sum_{j=1}^{k}\omega_j^2}{2N_{i}(t-1)}} 
     \geq \sum\limits_{j=1}^{k}\frac{\omega_j\mu_{1j}}{p_1} \label{eqn:neg1} \\
    &\geq  \frac{2r_{0}}{p_i}\sqrt{\frac{\alpha \ln n\sum_{j=1}^{k}\omega_j^2}{2N_{i}(t-1)}} + \sum\limits_{j=1}^{k}\frac{\omega_j\mu_{ij}}{p_i} \label{eqn:neg2} \\ 
    &\geq \sum\limits_{j=1}^{k}\frac{\omega_j\hat\mu_{ij}(t-1)}{p_i} + \frac{r_{0}}{p_i}\sqrt{\frac{\alpha \ln t\sum_{j=1}^{k}\omega_j^2}{2N_{i}(t-1)}},  \label{eqn:neg3}
\end{align}
\end{subequations}
where \eqref{eqn:neg1} follows by negating \eqref{eqn:UnderEstimate},  \eqref{eqn:neg2} follows by the definition of $\Delta_i$  and by using the negation of \eqref{eqn:GapBound}, and \eqref{eqn:neg3} follows by applying the negation of \eqref{eqn:OverEstimate} and $t \leq n$ in \eqref{eqn:neg2}. 
This contradicts the hypothesis that $I_t=i$.

Hence, if $I_t = i$ then $\{N_i(t) \leq u\}$ where we consider $u = \left\lceil \frac{2r_{0}^{2}\alpha \ln n\sum_{j=1}^{k}\omega_j^2}{p_i^2\Delta_{i}^{2}} \right\rceil$. We know $N_{i}(n)=\sum\limits_{t=1}^{n}\mathbb{I}\left\{I_{t}=i\right\}$. \text{Therefore, } 
\begin{align}
    & \EE{N_{i}(n)} = \EE{\sum\limits_{t=1}^{n}\mathbb{I}\left\{I_{t}=i\right\}} 
   \leq u + \sum\limits_{t=1}^{n}\Prob{I_{t}=i, N_{i}(t) > u}. \nonumber 
    \intertext{We know that if \eqref{eqn:GapBound} is false and $I_{t}=i$, then  \eqref{eqn:UnderEstimate} or \eqref{eqn:OverEstimate} must hold. So we get }\nonumber 
    & \EE{N_{i}(n)} \leq u + \sum\limits_{t=1}^{n}\Prob{\text{\eqref{eqn:UnderEstimate}) or \eqref{eqn:OverEstimate} must hold}}. \nonumber 
    \intertext{By applying the union bound, we get} 
    &\EE{N_{i}(n)} \leq u + \sum\limits_{t=1}^{n}[\Prob{ \eqref{eqn:UnderEstimate} \mbox{ holds}} + \Prob{\eqref{eqn:OverEstimate} \mbox{ holds}}], \label{eqn:bound2}
\end{align}
where
\begin{align}
     &\sum\limits_{t=1}^{n}\Prob{(\ref{eqn:UnderEstimate}) \mbox{ holds}} \leq \sum\limits_{t=1}^{n}\mathbb{P}\Biggl\{\exists s \in \{1,2,..,t\}; \sum\limits_{j=1}^{k}\omega_j\hat\mu_{1j}^{(s)} \nonumber \\
    &+ r_{0}\sqrt{\frac{\alpha \ln t\sum_{j=1}^{k}\omega_j^2}{2s}} \leq \sum\limits_{j=1}^{k}\omega_j\mu_{1j}\Biggr\} \nonumber \\
    & \leq \sum\limits_{t=1}^{n}\sum\limits_{s=1}^{t}\Prob{\sum\limits_{j=1}^{k}\omega_j\left(\mu_{1j} - \hat\mu_{1j}^{(s)}\right)\geq r_{0}\sqrt{\frac{\alpha \ln t\sum_{j=1}^{k}\omega_j^2}{2s}}}. \nonumber \\
    & \leq \sum\limits_{t=1}^{n}\sum\limits_{s=1}^{t-1}e^{-s \psi^{*}\left(r_{0}\sqrt{\frac{\alpha \ln t\sum_{j=1}^{k}\omega_j^2}{2s}}\right)} \qquad \text{By Lemma \ref{lem:ConcentrationBound1}} \nonumber\\
    & \leq \sum\limits_{t=1}^{n}\sum\limits_{s=1}^{t-1}e^{-\alpha \ln t}  \qquad \qquad \text{as $\psi^{*}(\epsilon) =\frac{2\epsilon^{2}}{r_{0}^{2}\sum_{j=1}^{k}\omega_j^2}$} \nonumber\\
    & \leq \sum\limits_{t=1}^{n}\sum\limits_{s=1}^{t}\frac{1}{t^{\alpha}} 
    \leq \sum\limits_{t=1}^{\infty}\frac{1}{t^{\alpha-1}}. \nonumber \\
    & \text{Considering $\alpha = 3$, we get } \sum\limits_{t=1}^{n}\Prob{\eqref{eqn:UnderEstimate} \mbox{ holds}} \leq \frac{\pi^{2}}{6}. \label{eqn:overEstimateBound} \\
    &\text{Similarly, we get \quad }
    \sum\limits_{t=1}^{n}\Prob{\eqref{eqn:OverEstimate} \mbox{ holds}} \leq \frac{\pi^{2}}{6}. \label{eqn:UnderEstimateBound} \\
    &\text{By inserting $u$, 
    \eqref{eqn:overEstimateBound}, and \eqref{eqn:UnderEstimateBound} into \eqref{eqn:bound2}, we get} \nonumber\\
    &\EE{N_{i}(n)} \leq \frac{6r_{0}^{2}\ln n\sum_{j=1}^{k}\omega_j^2}{p_i^2\Delta_{i}^{2}} +\left(\frac{\pi^{2}}{3} + 1\right). \label{eqn:SampleBound} \\
    &\text{By the Regret Decomposition Lemma \cite{lattimore2020bandit}, we have}\nonumber\\
    &\Regret_{n} = 
    \sum\limits_{i: \Delta_{i}>0}\Delta_{i}\EE{N_{i}(n)}. \label{eqn:regret decomposition}\\
    &\text{Inserting \eqref{eqn:SampleBound} into \eqref{eqn:regret decomposition} we get the desired bound in \eqref{regret:Thm1}}.\nonumber \qedhere
\end{align}
\end{proof}
\vspace{-3mm}